\documentstyle[amssymb, amsmath]{article}

\normalbaselines
\catcode `\@=11
\@addtoreset{equation}{section}

\def\section{\@startsection {section}{1}{\z@}{-3.5ex plus -1ex minus
     -.2ex}{2.3ex plus .2ex}{\normalsize\bf}}
\def\subsection{\@startsection{subsection}{2}{\z@}{-3.25ex plus -1ex minus
 -.2ex}{1.5ex plus .2ex}{\normalsize\bf}}

\def\thebibliography#1{\section*{References\markboth
  {REFERENCES}{REFERENCES}}\list
  {[\arabic{enumi}]}{\settowidth\labelwidth{[#1]}\leftmargin\labelwidth
  \advance\leftmargin\labelsep
  \usecounter{enumi}}
  \def\newblock{\hskip .11em plus .33em minus -.07em}
  \sloppy
  \sfcode`\.=1000\relax}
 
\catcode `\@=12
\def\K{{\Bbb K}}
\def\R {{\Bbb R }}
\def\C {{\Bbb C }}
\def\H {{\Bbb H }}

\def\UU{{\rm U}}
\def\U{{\rm U^\circ}}
\def\O{{\rm O}}
\def\Sp{{\rm Sp}}
\def\const{{\rm const}}
\def\B{{\rm B}}
\def\OO{{\rm SO}}
\def\SO{{\rm SO}}

\def\le{\leqslant}
\def\ge{\geqslant}

\def\Sti{{\rm Sti}}
\def\phi{{\varphi}}

\def\calo{{\frak O}}
\def\cA{{\cal A}}
\def\cB{{\cal B}}

\def\konets{\hfill$\square$}

\def\Re{{\rm Re}\,\,}

\newcommand{\matr}[1]{\left(\begin{array}{cc}#1\end{array}\right)}
\newcommand{\matra}[1]{\left(\begin{array}{ccc}#1\end{array}\right)}

\begin{document}

\vspace*{2.5cm}
\noindent
{ \bf HUA TYPE INTEGRALS OVER UNITARY GROUPS\\
 AND OVER PROJECTIVE LIMITS OF UNITARY GROUPS}\vspace{1.3cm}\\
\noindent
\hspace*{1in}
\begin{minipage}{13cm}
Yurii A.Neretin\footnote{supported by grants
 RFBR 98-01-00303 and NWO 047-008-009}\vspace{0.3cm}\\

  Institute of Theoretical and Experimental Physics,

  Bolshaya Cheremushkinskaya, 25,
  Moscow, 117259, Russia\\
 \&
Independent University of Moscow,
Bolshoi Vlas`evskii per., 11, Moscow, 121002, Russia\\
E-mail: neretin@main.mccme.rssi.ru
\end{minipage}

\vspace*{0.5cm}

\begin{abstract}
\noindent
We discuss some natural maps from a unitary group $\UU(n)$ to
a smaller group $\UU(n-m)$
(these maps are versions of the Liv\u{s}ic characteristic function).
 We calculate explicitly the
direct images of the Haar measure under some maps.
 We evaluate some matrix integrals
over classical groups and some symmetric spaces
(values of the integrals are products of $\Gamma$-functions).
These integrals generalize Hua Loo Keng integrals.
We  construct inverse limits of unitary groups
equipped with analogues of Haar measure and
 evaluate some integrals over these inverse limits.
\end{abstract}




 \vspace{22pt}

 \hfill {\it To memory of Sergei Kerov}

 \vspace{22pt}

Let  $\K$ be the real numbers  $\R$, the  complex numbers $\C$ or
the algebra of quaternions $\H$.
By $\UU(n,\K)=\O(n),\UU(n),\Sp(n)$ we denote the unitary group of the space $\K^n=\R^n,\C^n,\H^n$.
 We also will use the  notation
$$\U(n,\K):=\SO(n),\,\,\UU(n),\,\,\Sp(n)$$
for the connected component of the group $\UU(n,\K)$.

By $\sigma_n$ we denote
 the Haar measure on $\U(n,\K)$ normalized by the
condition  $\sigma_n(\U(n,\K))=1$.

Let $Q$ be a matrix over $\K$.  By $[Q]_p$ we denote
the left upper block of the matrix
$Q$ of  size $p\times p$. By $\{Q\}_p$ we denote the
right lower block of size $p\times p$.

 Let us represent a matrix $g\in\U(n,\K)$ as
a $(m+(n-m))\times (m+(n-m))$ block
 matrix $\matr{P&Q\\R&T}$. Consider the map
$$\Upsilon^m:  \matr{P&Q\\R&T}\mapsto T-R(1+P)^{-1}Q$$
(this map is defined almost everywhere on $\U(n,\K)$).

\smallskip

{\sc Proposition 0.1}(\cite{Nerr})
  a){\it  $\Upsilon^m$ maps $\U(n,\K)$ to $\U(n-m,\K)$.}

  b){\it $\Upsilon^k\circ\Upsilon^m=\Upsilon^{k+m}$.  }

  c){\it $S=\frac{g-1}{g+1}$ implies $\{S\}_p
=\frac{\Upsilon^{n-p}(g)-1}{\Upsilon^{n-p}(g)+1}$
                                            }

\smallskip

{\sc Remark.} It is clear that
 the map $\Upsilon^m$ is not a homomorphism
$\U(n,\K) \to \U(n-m,\K)$.
It is, however, a morphism of symmetric spaces in
a sense explained in \cite{NerKS}.                   \konets

\smallskip

{\sc Remark.}
Note that $\Upsilon^m$ is a value of the Liv\u{s}ic characteristic function
$$\chi(\lambda)=T+\lambda R(1-\lambda P)^{-1}Q$$
at  $\lambda=-1$ (see \cite{Liv}) and Proposition 0.1 is quite a standard claim
from this point of view.  Characteristic functions were widely exploited
for studies of spectral  properties of an individual operator
(see \cite{Nik}).
Nevertheless, it seems that until \cite{Nerr}
 they have never been used in  analysis on unitary groups.  \konets

\smallskip

{\sc Remark.}
The statement c) means the following. For a matrix $g\in\U (\K)$
consider its Cayley transform $(1-g)^{-1}(1+g)$. Then we
consider the
inverse Cayley transform of the right lower block of $(1-g)^{-1}(1+g)$.
The result coincides with the application of the map $\Upsilon$ to $g$.
                                        \konets

\smallskip

In Section 1 we investigate some properties of
the maps $\Upsilon^m$.
We construct maps from the groups $\U(n,\K)$   to some spaces
(groups of lower dimension, cubes, products of matrix balls, etc.)
and calculate explicitly the images of the Haar measure under these maps.

We also consider the following measures on the groups $\U(n,\K)$
\begin{equation}
\prod_{k=1}^{n}
 |\det(1+[g]_{n-k+1})|^{\lambda_k-\lambda_{k-1}} d\sigma_n(g).
\end{equation}
(we assume $\lambda_0=0$).
We show that the image of such a measure with respect to
the map $\Upsilon^1$ coincides with
the measure on $\U(n-1,\K)$ given by the formula
$$
C(\lambda)\prod_{k=1}^{n-1}
|\det(1+[h]_{n-k})|^{\lambda_k-\lambda_{k-1}} d\sigma_{n-1}(h),
$$
where the value  $C(\lambda)$ is explicitly evaluated  (Theorem 1.5).

This observation makes it possible
to obtain pleasant explicit formulas (2.1)--(2.3) for the
 integrals of the functions (0.1). For instance, in the case $\K=\C$
we obtain
$$\int_{\UU(n)}\prod_{k=1}^{n}\left\{
\det(1+[g]_{n-k+1})^{\lambda_k -\lambda_{k-1} }
\overline{\det(1+[g]_{n-k+1})}^{\mu_k-\mu_{k-1}}\right\}\,d\sigma_n(g)
             =
\prod_{k=1}^{n} \frac{\Gamma(k)\Gamma(k+\lambda_k+\mu_k)}
{\Gamma(k+\lambda_k)\Gamma(k+\mu_k)}                               .
              $$
In the case
 $\lambda_1=\dots=\lambda_{n}=\overline{\mu_1}=\dots=\overline{\mu_{n}}$
we obtain one of Hua Loo Keng's integrals (\cite{Hua}, Chapter 2).
In Section 2 we also discuss other matrix integrals.

In Section 3 we construct inverse limits of the unitary
groups (virtual unitary groups) and give some remarks on these
limits.
 Virtual unitary groups
  are  close to Pickrell's virtual Grassmannian
(see \cite{Pic}, see also Shimomura's paper \cite{Shim})
and to Kerov--Olshanski--Vershik's virtual permutations
(\cite{KOV}). Explicit
Plancherel formula for the virtual unitary groups is obtained
in  recent works of Borodin and Olshanski \cite{this},
\cite{BO3}
and \cite{Ols2}. 

 Another application of our integrals is a separation
of spectra in analysis of Berezin kernels (\cite{NO}, \cite{Nerr}).

 I thank G.I.Olshanski for discussions of this subject.
 I also thank the referee of the paper for his comments.

\section{\hspace{-4mm}.\hspace{2mm} MAPS  $\Upsilon^m$
  AND PROJECTIONS OF MEASURES}

\subsection{\hspace{-5mm}.\hspace{2mm} Proof of Proposition 0.1}
First we prove c).
We use the following Frobenius formula (see  \cite{Gan}, Section II.5)
 for the inverse of a block matrix
\begin{eqnarray}
\matr{A&B\\C&D}^{-1}=
\matr{A^{-1}+A^{-1}B(D-CA^{-1}B)^{-1}C A^{-1}&
- A^{-1}B(D-CA^{-1}B)^{-1}
\\-(D-CA^{-1}B)^{-1} C A^{-1} & (D-CA^{-1}B)^{-1} }   \\
=\matr{(A-BD^{-1}C)^{-1}    &
-(A-BD^{-1}C)^{-1} BD^{-1}\\ - D^{-1} C(A-BD^{-1}C)^{-1}   &
D^{-1}+ D^{-1} C(A-BD^{-1}C)^{-1}  BD^{-1} } .
\end{eqnarray}
We have $S=-1+2(1+g)^{-1}$  and  formula (1.1) implies
$$\{S\}_p= -1+2(1+T-R(1+P)^{-1}Q))^{-1}=-1+2(1+\Upsilon^{n-p}(g))^{-1}.$$

   The statement b) is a  consequence of c).

   The condition $g\in \U(n,\K)$ is equivalent to the condition
$S+S^*=0$. This implies  a).\konets

\smallskip
{\sc Remark.} A proof of Proposition 0.1 which does not require any  calculations is contained in \cite{Nerr}.

\subsection{\hspace{-5mm}.\hspace{2mm} Projection of the Haar Measures}

{\sc Lemma 1.1.} {\it
 Let $A,B\in\U(n-m,\K)$. Then}
$$\Upsilon^m\left[\matr{1&0\\0&A}g\matr{1&0\\0&B}\right]=
A\Upsilon^m(g)B.$$

\smallskip

{\sc Proof.} Obvious.  \konets

\smallskip

{\sc Corollary 1.2.} {\it  The image of the
probability Haar measure $\sigma_n$ on $\U(n,\K)$
under the map $\Upsilon^m: \U(n,\K)\to\U(n-m,\K)$
 is the Haar measure $\sigma_{n-m}$.}

\smallskip

{\sc Proof.} Indeed, the image is the
probability $\U(n-m,\K)
\times\U(n-m,\K)$-invariant measure on $\U(n-m,\K)$. \konets

\smallskip

By $\B_n=\B_n(\K)$ we denote a set of all $n\times n$-matrices $Z$
over $\K$ satisfying
the condition $\|Z\|<1$ (here $\|\cdot\|$ denotes the norm of an operator
in an Euclidean space
$\K^n$).

Consider a map
$$\xi_m:\U(n,\K)\to  \U(n-m,\K)\times \B_m$$
defined by
$$\xi_m (g)=(\Upsilon^m(g),[g]_m).$$
Recall that the map $\Upsilon^m$ is defined
almost everywhere.

\smallskip

{\sc Theorem 1.3.} {\it    Suppose $n\ge 2m$.
Let
 $$\tau=(n-2m+1)\dim\K/\,2.$$
Then the image of the  Haar measure $\sigma_{n}(g)$
 with respect to the map $\xi_m$ is given by the formula
\begin{equation}
\const\cdot \det(1-Z^*Z)^{\tau-1}\,dZ\,d\sigma_{n-m}(h),
\end{equation}
where  $ Z\in \B_m, \,h\in\U(n-m,\K)$, and
 $dZ$ is the Lebesgue measure on $\B_m$.}

\smallskip

{\sc Remark.} Let us recall the definition of the quaternionic determinant.
 Let $g$
be a quaternionic operator $\H^n\to\H^n$. We can consider $g$ as
an operator
$g_\R:\R^{4n}\to\R^{4n}$. Then
$$\det g:=\sqrt[4]{\det g_\R}.$$
The quaternionic linear group ${\rm GL}(n,\H)$
is connected and hence the determinant under the root
is positive. If $g$ is a diagonal matrix
with values $a_1$, \dots, $a_m$ on the diagonal,
then $\det(g)=\prod |a_j|$.\konets

\smallskip

{\sc Remark.} For a matrix $X$ satisfying $\|X\|<1$,
the power
$$(1+X)^\lambda:=\sum_{l=0}^\infty
 \frac{\lambda(\lambda-1)\dots(\lambda-l+1)}{l!}X^l
 $$
 is well defined. Hence, in (1.3) the expression
 $(1-Z^*Z)^{\tau-1}$ is well defined, hence its determinant
 also is well defined. \konets

\smallskip

{\sc Proof.} Denote by $\nu$ the image of the Haar measure
on $\U(n,\K)$
under the map $\xi_m$.
 Let $A,B\in\U(n-m,\K)$. By Lemma 1.1,
$$\xi_m\left[\matr{1&0\\0&A}g\matr{1&0\\0&B}\right]=
(A\Upsilon^m(g)B,[g]_m).$$
Hence, the measure $\nu$ on  $\U(n-m,\K)\times \B_m$ is
 invariant with respect
to the transformations $(h,Z)\mapsto (AhB,Z)$. Thus, $\nu$ has the form
$\phi(Z)\,dZ\,d\sigma_{n-m}(h)$,
where $\phi(Z)$ is a function on $B_m$.
We want to calculate $\phi(Z)$. For this purpose,
 we project  the measure $\nu$ from  $ \U(n-m,\K)\times\B_m $ to $\B_m$.
Obviously, the image of $\nu$ under
 this projection has a form $\const\cdot\phi(Z)\,dZ$.

Consider the simplex $\Sigma_m\subset\R^m$ defined by the inequalities
$1\ge r_1\ge \dots\ge r_m\ge 0$. Each point $Z$ of the matrix ball $\B_m$
can be represented in the form
$$Z=q_1\cdot\matra{r_1&&\\&r_2&\\&&\ddots}\cdot q_2;\quad  \mbox{where }
q_1,q_2\in \U(m,\K), \quad (r_1,r_2,\dots)\in\Sigma_m.$$
Obviously, the numbers $r_j$ are uniquely defined by the matrix $Z$.

To evaluate the density $\phi(Z)=\frac{\phi(Z)\,dZ}{dZ}$ we project
both measures $\phi(Z)\,dZ$, $dZ$ to the simplex $\Sigma_m$.
 The projection of the measure
$\phi(Z)\,dZ$ to $\Sigma_m$  coincides with the projection of the Haar measure
from $\U(n,\K)$
to $\Sigma_m$. The latter projection
is the  radial part of the
Haar measure on $\U(n,\K)$ with respect to the symmetric
subgroup $\U(m,\K)\times\U(n-m,\K)$.
An explicit formula for the  radial part
(see \cite{Hel}, X.1) is given by:
\begin{equation}  \const\cdot
  \prod_{1\le i\le m} (1-r_i^2)^{((n-2m+1)\dim\K-2)/2}
  \Bigl\{\prod_{1\le i<j\le m}(r_i^2-r_j^2)^{\dim\K} \prod_{1\le i\le m}
   r_i^{\dim\K-1}
 \prod_{1\le i\le m} dr_i\Bigr\}
.\end{equation}

It is easy to calculate the projection of the Lebesgue measure
$dZ$ to $\Sigma_m$ (see calculations
of this type in  \cite{Hua}, chapters 2-3).
 This projection  coincides with the expression in the curly
brackets in (1.4).

It remains to observe that the function $\det(1-Z^*Z)$ is
 $\U(m,\K)\times\U(m,\K)$-invariant, and its restriction to
 the simplex $\Sigma_m$
equals $\prod(1-r_i^2)$.           \konets

\smallskip

The measure (1.3) have to be a probability
measure, and hence the constant in (1.3)
is inverse to the following value  $c_\K^{(m)} (\tau)$.

\smallskip

{\sc Lemma 1.4.} {\it For $\tau>0$,}
$$
c_\K^{(m)} (\tau):=\int\limits_{B_m(\K)} \det (1-Z^*Z)^{\tau-1}\,dZ=
\pi^{m^2\dim\K\,/2} \prod_{j=1}^{m} \frac{\Gamma(\tau+(j-1)\dim\K\,/2)}
                           {\Gamma(\tau+(m+j-1)\dim\K\,/2)}.
$$

In this formula, we use the simplest normalization of
the Lebesgue measure $dZ$.
If $\K=\R$ and $z_{\alpha\beta}$ are the matrix elements of $Z$, then
$dZ:=  \prod dz_{\alpha\beta}$. For $\K=\C$, we represent
matrix elements in the form
          $z_{\alpha\beta}= u_{\alpha\beta}+i v_{\alpha\beta}$, and
          assume $dZ:= \prod  du_{\alpha\beta}dv_{\alpha\beta}$.
For $\K=\H$, we write
$z_{\alpha\beta}=
 u_{\alpha\beta}+{\bold i} v_{\alpha\beta}+
{\bold j} w_{\alpha\beta}
+{\bold k}  h_{\alpha\beta}$,
and assume
$dZ=
 du_{\alpha\beta}\, dv_{\alpha\beta}\, dw_{\alpha\beta}
\,dh_{\alpha\beta}
$.

\smallskip

{\sc Remark.}
In particular,   in the case $m=1$
the normalizing constants in Theorem 1.3
are

a) for $\K=\R$\,:\,\,\,\, $\pi^{-1/2}\Gamma(n/2)/\Gamma((n-1)/2)$;

b) for $\K=\C$\,: \,\,\,\, $(n-1)/\pi$;

c) for $\K=\H$\,: \,\,\,\, $(2n-2)(2n-1)/\pi^2$.  \konets

{\sc Proof.} In principle, these integrals
were evaluated by
 Hua Loo Keng. But he considered only
 case $\K=\C$. Obviously, his method
also is quite valid in two other cases.
To avoid calculations, we  give
 a reduction to the Selberg integral.

Denote $\delta=\dim\K$. By (1.4),
 $$c_\K^{(m)} (\tau)=C(m,\delta)\cdot \int\limits_0^1\dots\int\limits_0^1
  \prod_{1\le i\le m} (1-r_i^2)^{\tau-1}
  \prod_{1\le i<j\le m}(r_i^2-r_j^2)^{\delta} \prod_{1\le i\le m}
   r_i^{\delta-1}
 \prod_{1\le i\le m} dr_i
 ,$$
where $C(m,\delta)$ is a constant which does not depend
on $\tau$. We substitute
$x_j=r_j^2$
and apply the Selberg integral (see \cite{AAR})
\begin{multline*}
 \int\limits_0^1\dots\int\limits_0^1
 \prod_{1\le i< j\le n} \bigl|x_i-x_j\bigr|^{2\gamma}
 \prod_{1\le j\le n} \bigl\{ x_j^{\alpha-1}(1-x_j)^{\beta-1}\bigr\}
  \prod_{1\le j\le n} dx_j =\\  =
  \prod_{j=1}^n \frac
     {\Gamma(\alpha+(j-1)\gamma)\,
     \Gamma(\beta+(j-1)\gamma)\,\Gamma(1+j\gamma)}
      {\Gamma(\alpha+\beta+(n+j-2)\gamma)\,\Gamma(1+\gamma)}.
\end{multline*}
This gives
$$ c_\K^{(m)} (\tau)=
\frac 1{2^m}  C(m,\delta)
\prod_{j=1}^m
\frac{\Gamma(\tau+(j-1)\delta/2)\, \Gamma(j\delta/2)\,  \Gamma(1+j\delta/2)}
     {\Gamma(\tau+(m+j-1)\delta/2)\, \Gamma(1+\delta/2)}.
$$

To find the constant, we obtain the asymptotics
of   $c_\K^{(m)} (\tau)$ as $\tau\to+\infty$ in two ways.
First, we apply the formula
$$\Gamma(a+x)/\Gamma(b+x)\sim x^{(a-b)},\qquad x\to +\infty,$$
 and obtain
 $$c_\K^{(m)} (\tau)\sim
 \tau^{-m^2\delta/2}\cdot  \frac 1{2^m}  C(m,\delta)
\prod_{j=1}^m \frac{
 \Gamma(j\delta/2)\,  \Gamma(1+j\delta/2)}
      { \Gamma(1+\delta/2)},\qquad \tau\to+\infty.
 $$
 Applying the Laplace method (see, for instance, \cite{Fed}),
 we obtain
\begin{multline*}
  c_\K^{(m)} (\tau):=\int\limits_{\|Z\|<1} \det (1-Z^*Z)^{\tau-1}\,dZ=
  \tau^{-m^2\delta/2}
  \int
  \limits_{\|Z\|<\sqrt \tau} \det (1-\tfrac 1\tau Z^*Z)^{\tau-1}\,dZ\sim\\
  \sim
  \tau^{-m^2\delta/2} \int \exp\bigl( -{\rm tr}\, Z^*Z\bigr)\,dZ=
  \tau^{-m^2\delta/2}  \pi^{m^2\delta/2}; \qquad  \tau\to+\infty;
\end{multline*}
the last integral is taken over the whole space of $m\times m$
 matrices.
Compairing the asymptotics we obtain the explicit expression
for $C(m,\delta)$.
\konets

\subsection{\hspace{-5mm}.\hspace{2mm} Projection of
an  Orthogonal Group to a Cube}

Assume  $\K=\R$ (other cases are similar).
 Let $m=1$ in the notation of Subsection 1.2.
In this case, the 'ball' $\B_1$
is the segment $[-1,1]$.
Consider the iterations of the map $\xi_1$:
$$\OO(n)\to  \OO(n-1)\times  [-1,1]
\to  \OO(n-2)\times[-1,1] \times  [-1,1]\to\dots$$
 We obtain a map $\Theta$
 (defined almost everywhere)
 from $\OO(n) $ to the cube
$[-1,1]^{n-1}=[-1,1]\times\dots\times[-1,1]$
given by the formula
$$(x_2,x_3,\dots,x_n)=\bigl
([\Upsilon^{n-2}(g)]_1, \dots,
[\Upsilon^1(g)]_1,[g]_1\bigr).$$

{\sc Remark.} The map
$\Upsilon^{n-1}$ maps $\SO(n)$ to $\SO(1)$.
The latter group is a singleton and hence we can omit
$\Upsilon^{n-1}$ from the formula. In the case $\K=\C$,
we obtain a map from $\UU(n)$ to $S^1\times\B_1(\C)^{n-1}$
where $B_1(\C)$ is the disk $|z|\le 1$ in $\C$
 and $S^1$ is
the circle $|z|=1$. In the case $\K=\H$,
we obtain a map $\Sp(n)\to S^3\times B_1(\H)^{n-1}$,
there $S^3$ is the sphere $|z|=1$ in $\H\simeq\R^4$ and
$B_1(\H)$ is the ball in $\H$.\konets

\smallskip

By Theorem 1.3, the image of the
probability Haar measure under our map $\Theta$
 equals
$$d\mu(x_2,\dots,x_n)=\frac{\Gamma(n/2)}{\pi^{n/2}}
\prod_{j=2}^{n} (1-x_j^2)^{(j-3)/2}
\prod_{j=2}^{n}dx_j.$$
Consider a function $f$ depending of $n-1$ variables. Then we have
\begin{equation}
\int_{\OO(n)}
f([\Upsilon^{n-2}(g)]_1, \dots,[\Upsilon^1(g)]_1,[g]_1)\,d\sigma_n(g)=
\int_{[-1,1]^{n-1}} f(x_2,\dots,x_n)  d\mu(x_2,\dots,x_n).
\end{equation}

\subsection{\hspace{-5mm}.\hspace{2mm} Projections of Orthogonal Groups to
Products of Matrix Balls}
Let $n=p_1+\dots+ p_\alpha+q$ and
$p_j\le p_{j+1}+\dots+p_\alpha+q$ for all $j$. Consider a map
$$\Theta:\OO(n)\to\OO(q)\times\B_{p_\alpha}\times\dots\times\B_{p_1}$$
given by the formula
$$\Theta(g)=(\Upsilon^{p_1+\dots+p_\alpha}(g),
[ \Upsilon^{p_1+\dots+p_{\alpha-1}}(g)]_{ p_\alpha},\dots,
 [\Upsilon^{p_1}(g)]_{p_2},[g]_{p_1})$$

By Theorem 1.3, the image of the Haar measure
on $\OO(n)$ under the map
$\Theta $ is
$$\const\cdot
\prod_{j=1}^{\alpha}
 \det(1-Z_j^*Z_j)^{(p_{j+1}+\dots+p_\alpha+q-p_j-1)/2}
  \prod_{j=1}^{\alpha} dZ_j  \,\,
d\sigma_q(h),$$
where
$h\in\OO(q)$, $Z_j\in\B_{p_j}$ and the constant
is  the product of constants evaluated in Lemma 4:
$$\prod_{j=1}^\alpha c^{p_j}_\R
\bigl(p_{j+1}+\dots+p_\alpha-p_j+1)/2\bigr).$$

\subsection{\hspace{-5mm}.\hspace{2mm} Multiplicativity}
{\sc Proposition 1.4.}{ \it Let $g\in\U(n,\K)$ and let $m<p\le n$. Then}
$$\det(1+[g]_p)=\det(1+[g]_m)\det(1+[\Upsilon^m(g)]_{p-m})$$

\smallskip

{\sc Proof.} Let us represent $g$  as  an
$(m+(p-m)+(n-p)) \times (m+(p-m)+(n-p)) $ block matrix  :
$$g=\matra{P&Q_1&Q_2\\R_1&T_{11}&T_{12} \\R_2&T_{21}&T_{22}}. $$

By the formula
$$\det\begin{pmatrix}A&B\\C&D\end{pmatrix}=
   \det(A)\det(D-CA^{-1}B).$$
for the determinant of  a block matrix
(\cite{Gan}, Section II.5), we obtain
$$\det\left[1+\matr{P&Q_1\\R_1&T_{11}}\right]=
\det(1+P)\det(1+T_{11}-R_1(1+P)^{-1}Q_1).$$
On the other hand, we have
$$\Upsilon^m(g)=
\matr{T_{11}-R_1(1+P)^{-1}Q_1&T_{12}-R_1(1+P)^{-1}Q_2  \\
      T_{21}-R_2(1+P)^{-1}Q_1  & T_{22}-R_2(1+P)^{-1}Q_2}.$$
 and the statement obviously follows.
                            \konets

\smallskip

{\sc Remark.}
Proposition 1.5 gives an expression of the coordinates $x_j$
on the cube (see  1.3)
$$
1+x_{n-m}=1+[\Upsilon^m(g)]_1=\frac{\det(1+[g]_{m+1})} {\det(1+[g]_{m})}.
$$
Also we obtain we identity
$$
\det(1+[g]_m)=\prod_{j=1}^m (1+[\Upsilon^{j-1}(g)]_1).
$$

\subsection{\hspace{-5mm}.\hspace{2mm} A Consistent System of Measures}
Consider $\lambda_1,\dots,\lambda_{n},\mu_1,\dots,\mu_{n}\in\C$.
Assume $\lambda_0=\mu_0=0$.

\smallskip

{\sc Theorem 1.6.}{\it

{\rm a)} Let $\K=\R$. Assume $\Re\lambda_n>-(n-1)/2$.
 Consider the measure on $\OO(n)$
  given by the formula
\begin{equation}
\prod_{k=1}^{n}
\det(1+[g]_{n-k+1})^{\lambda_k-\lambda_{k-1}}\,d\sigma_n(g).
\end{equation}
Then its image under the map $\Upsilon^1$ is
$$2^{\lambda_n}\frac{\Gamma(n-1)\Gamma(\lambda_n+(n-1)/2)}
       {\Gamma((n-1)/2)\Gamma(\lambda_n+n-1)}
\prod_{k=1}^{n-1}\det(1+[h]_{n-k})^{\lambda_k-\lambda_{k-1}}\,
 d\sigma_{n-1}(h).$$

{\rm b)} Let  $\K=\C$. Assume $\Re(\lambda_n+\mu_n)>-n$.
 Consider the {\rm(} complex-valued{\rm)}
  measure on    $\UU(n)$ given by the formula
\begin{equation}
\prod_{k=1}^{n}
\det(1+[g]_{n-k+1})^{\lambda_k-\lambda_{k-1}}
\overline{\det(1+[g]_{n-k+1})}^{\mu_k-\mu_{k-1}}\,d\sigma_n(g)
.\end{equation}
Then its image under the map $\Upsilon^1$ is
$$\frac{\Gamma(n)\Gamma(n+\lambda_n+\mu_n)}
{\Gamma(n+\lambda_n)\Gamma(n+\mu_n)}
\prod_{k=1}^{n-1}
\det(1+[h]_{n-k})^{\lambda_k-\lambda_{k-1}}\overline
{\det(1+[h]_{n-k})}^{\mu_k-\mu_{k-1}}\,d\sigma_{n-1}(h).$$

{\rm c)} Let $\K=\H$. Assume $\Re\lambda_n>-2n-1$.
 Consider the measure on      $\Sp(n)$ given by the formula
\begin{equation}
\prod_{k=1}^{n}\det(1+[g]_{n-k+1})^{\lambda_k-\lambda_{k-1}}
   \,d\sigma_n(g).
\end{equation}
Then its image under the map $\Upsilon^1$ is
$$\frac{\Gamma(2n)\Gamma(2n+\lambda_n+1)}
{\Gamma(2n+\lambda_n/2)\Gamma(2n+\lambda_n/2+1)}
\prod_{k=1}^{n-1}\det(1+[h]_{n-k})^{\lambda_k-\lambda_{k-1}}
\,d\sigma_{n-1}(h).$$}

\smallskip

{\sc Remark.} The measure (1.7) is a positive real-valued
 measure iff
 $\lambda_j=\overline\mu_j$ for all $j$.     \konets


\subsection{\hspace{-5mm}.\hspace{2mm} Proof of Theorem 1.6 for $\K=\R$}

We apply Theorem 1.3.
In our case
the matrix ball $\B_1$ is the segment $[-1,1]$ and  $[g]_1=g_{11}$
is the left upper matrix element of $g\in \SO(n)$.
 By Proposition 1.5,  we have
\begin{eqnarray*}
\prod_{k=1}^{n}\det(1+[g]_{n-k+1})^{\lambda_k-\lambda_{k-1}}=
\left\{\prod_{k=1}^{n}(1+[g]_{1})^{\lambda_k-\lambda_{k-1}}\right\} \times
\left\{\prod_{k=1}^{n-1}
\det(1+[\Upsilon^1(g)]_{n-k})^{\lambda_k-\lambda_{k-1}}\right\}
.\end{eqnarray*}
Hence, the projection of the measure (1.6) to $\B_1\times \SO(n-1)$ is
\begin{align}
\frac{\Gamma(n/2)}{\pi^{1/2}\Gamma((n-1)/2)}
\left\{(1+x)^{\lambda_n}(1-x^2)^{(n-3)/2}dx\right\} \times
\qquad \qquad \qquad \qquad \qquad  \qquad
 \qquad \qquad \qquad
\\
\times  \left\{\prod_{k=1}^{n-1}
\det(1+[h]_{n-k})^{\lambda_k-\lambda_{k-1}}d\sigma_{n-1}(h)\right\},
\qquad \text{where $x\in[-1,1]$, $h\in \SO(n-1)$,}
 \end{align}
i.e., this projection is a product-measure.
Thus, the projection of the Haar measure to
 $\SO(n-1)$
is given by  formula
(1.10) up to a factor depending on $\lambda_n$.
The factor  is
(to simplify the formula we write
$\lambda$ instead of $\lambda_n$):
\begin{align}
 &\frac{\Gamma(n/2)}{\pi^{1/2}\Gamma((n-1)/2)}
 \int_{-1}^1(1-x^2)^{(n-3)/2}(1+x)^\lambda dx=
\,\,\,\,\,\,\,\,\,\,\,\,\,\,\,\,\,\,\,\,
\\= &
\frac{\Gamma(n/2)}{\pi^{1/2}\Gamma((n-1)/2)} \cdot
2^{\lambda+n-2}B\left(\lambda+\tfrac{n-1}{2},\tfrac{n-1}{2}  \right)=
2^{\lambda+n-2}
 \frac{\Gamma(n/2)\Gamma(\lambda+(n-1)/2)}
{\pi^{1/2}\Gamma(\lambda+n-1)} \nonumber
.\end{align}
Applying the duplication formula for $\Gamma$, we obtain
the required statement

\subsection{\hspace{-5mm}.\hspace{2mm}
 Proof of Theorem 1.6 for $\K=\C$}

The proof is similar. In the case $\K=\C$, the integral
 (1.11) is replaced by the integral
\begin{equation}
J_n(\lambda,\mu)=
\int\!\!\int_{|p|\le 1}(1+p)^\lambda(1+\overline p)^\mu(1-|p|^2)^{n-2}
\,\frac1idpd\overline p
.\end{equation}

After the substitution  $p=re^{i\phi},   \overline p=re^{-i\phi}$ we obtain
\begin{equation}
J_n(\lambda,\mu)=
\int^1_0dr\int_0^{2\pi}(1+re^{i\phi})^\lambda(1+re^{-i\phi})^\mu
(1-r^2)^{n-2}r\,d\phi.
\end{equation}
Expanding two first factors of the integrand into the Taylor series, we get
$$J_n(\lambda,\mu)=\int^1_0dr\int_0^{2\pi}
\left(\sum_{k=0}^\infty\frac{(-\lambda)_k}{k!}(-r)^ke^{ik\phi}\right)
\left(\sum_{k=0}^\infty\frac{(-\mu)_k}{k!}(-r)^ke^{-ik\phi}\right)
(1-r^2)^{n-2}r\,d\phi.$$
The integration over $\phi$ gives
$$\pi \int_0^1    \sum_{k=0}^\infty
\left\{\frac{(-\lambda)_k(-\mu)_k}{k!k!}
r^{2k}(1-r^2)^{n-2}\right\}dr^2.$$
Finally integrating in $r$, we get
\begin{equation}J_n(\lambda,\mu)=\pi     \sum_{k=0}^\infty \frac{(-\lambda)_k(-\mu)_k}{k!k!}
 \int_0^1 x^{k}(1-x)^{n-2}\,dx=\end{equation}
$$=\pi     \sum_{k=0}^\infty \frac{(-\lambda)_k(-\mu)_k(n-2)!}{k!(k+n-1)!}=
\frac{\pi }{n-1} F(-\lambda,-\mu;n;1).$$

The application of Gauss' formula
\begin{equation}
F(a,b;c;1)=\frac{\Gamma(c)\Gamma(c-a-b)}{\Gamma(c-a)\Gamma(c-b)}
\end{equation}
yields
$$J_n(\lambda,\mu)=
\frac{\pi }{n-1}
\frac{\Gamma(n)\Gamma(n+\lambda+\mu)}{\Gamma(n+\lambda)\Gamma(n+\mu)}.$$

After multiplication by $(n-1)/\pi$
(see the normalization constants after Lemma 1.4) we obtain the
required statement.

\subsection{\hspace{-5mm}.\hspace{2mm}
 Proof of Theorem 1.6 for $\K=\H$}

In the case  $\K=\H$, we obtain the integral
\begin{multline}
J_n(s)= \\
=\int\! \int\! \int\! \int_{h_1^2+h_2^2+ h_3^2+ h_4^2 \le 1}
((1+h_1)^2+ h_2^2+ h_3^2+ h_4^2)^{\lambda/2}(1-h_1^2-h_2^2- h_3^2- h_4^2)^{2n-3}
\,dh_1 \,dh_2 \,dh_3  \,dh_4
.\end{multline}
Passing to the spherical coordinates with respect to the variables
    $h_2, h_3 ,h_4$,
we obtain
$$J_n(\lambda)= 4\pi  \int\! \int
((1+h_1)^2+\rho^2)^{\lambda/2} (1-h_1^2-\rho^2)^{2n-3}
\rho^2\,dh_1\,d\rho,$$
where the domain of integration is $h_1^2+\rho^2 \le 1,\rho\ge0$.
We can change the integral to the integral over the whole circle
$h_1^2+\rho^2 \le 1$
(of course, we must write $1/2$  in front of the integral).
Passing to the polar coordinates
$h_1=r\cos\phi,\rho=r\sin\phi$, we obtain
$$2\pi \int_0^1dr\int_0^{2\pi}
(1+re^{i\phi})^{\lambda/2}(1+re^{-i\phi})^{\lambda/2}(1-r^2)^{2n-3}
r^3\cdot\left(- \frac14\right)
\left\{e^{2i\phi}-2+ e^{-2i\phi}\right\}\,d\phi.$$

Removing the curly brackets, we obtain a sum of 3 integrals.
The first and the third integrals coincide.  We obtain
\begin{eqnarray*}
J_n(\lambda)=\pi  \int_0^1dr\int_0^{2\pi}
(1+re^{i\phi})^{\lambda/2}(1+re^{-i\phi})^{\lambda/2}
(1-r^2)^{2n-3}r^3\,d\phi+\\
-\pi  \int_0^1dr\int_0^{2\pi}
(1+re^{i\phi})^{\lambda/2}
(1+re^{-i\phi})^{\lambda/2}(1-r^2)^{2n-3}r^3e^{2i\phi}
\,d\phi
.\end{eqnarray*}
Repeating the calculations (1.13) -- (1.14) for each integral, we get
\begin{eqnarray*}
J_n(\lambda)=\pi^2 \left[-\sum\limits_{k=0}^\infty
\frac{(-\lambda/2)_{k+2}(-\lambda/2)_{k}   (2n-1)!}
                  {k!(2n+k)!}+
\sum\limits_{k=0}^\infty (k+1)\frac{(-\lambda/2)_k(-\lambda/2)_k(2n-3)!}{k!(2n+k-1)!}\right]=\\
=\pi^2 \Biggl[-\frac{(-\lambda/2)(-\lambda/2+1)}{2n(2n-1)(2n-2)}
  F(-\lambda/2+2,-\lambda/2;2n+1;1)
+\frac1{(2n-1)(2n-2)}F(-\lambda/2,-\lambda/2;2n;1)+\\
+\frac{(-\lambda/2)^2}{2n(2n-1)(2n-2)}
F(-\lambda/2+1,-\lambda/2+1;2n+1;1)\biggr]
.\end{eqnarray*}
Applying Gauss' formula (1.15),
we obtain
\begin{multline*}
\pi^2\Biggl[ -\frac{(-\lambda/2)(-\lambda/2+1)}{2n(2n-1)(2n-2)}
\cdot\frac{(2n)!\Gamma(2n+\lambda-1)}
  {\Gamma(2n+\lambda/2-1) \Gamma(2n+\lambda/2+1)}+
   \frac1{(2n-1)(2n-2)}
   \cdot\frac{(2n-1)!\Gamma(2n+\lambda)}
     {\Gamma^2(2n+\lambda/2)}+         \\
 +\frac{(-\lambda/2)^2}{2n(2n-1)(2n-2)}
    \cdot\frac{(2n)!\Gamma(2n+\lambda-1)}
      {\Gamma^2(2n+\lambda/2)} \Biggr]=\\
= \pi^2
 \cdot\frac{(2n-3)!\Gamma(2n+\lambda-1)}
   {\Gamma(2n+\lambda/2-1) \Gamma(2n+\lambda/2)}
   \Biggl[-\frac{(-\lambda/2)(-\lambda/2+1)}
          {2n+\lambda/2}+\frac{2n+\lambda-1}{2n+\lambda/2-1}
    \frac{(-\lambda/2)^2}{2n+\lambda/2-1} \Biggr]
 .   \end{multline*}
Finally,
we obtain
$$J_n(\lambda  )=\pi^2 \frac{(2n-3)!\Gamma(2n+\lambda   +1)}
    {\Gamma(2n+\lambda   /2)  \Gamma(2n+\lambda   /2+1)}
    =
 \frac{ \pi^2}{(2n-1)(2n-2)}\,\cdot\, \frac{(2n-1)!\Gamma(2n+\lambda   +1)}
    {\Gamma(2n+\lambda   /2)  \Gamma(2n+\lambda   /2+1)}
    \,.
  $$

Now we have to repeat all calculations for the case $n=1$.
In the case   $\K=\C$ we have an integral over the circle $|p|=1$, in the case $\K=\H$ we obtain an integral over the sphere $h_1^2+h_2^2+h_3^2+h_4^2=1$.
We omit these calculations.

\subsection{\hspace{-5mm}.\hspace{2mm}
\bf
 Another representation of the densities.}

For definiteness, consider the case $\K=\R$.
Multiplicativity theorem 1.5
shows that  the density  (1.6) can be represented in the form
\begin{equation}
 \prod_{k=1}^{n}
 \det(1+[g]_{n-k+1})^{\lambda_k-\lambda_{k-1}}\cdot d\sigma_n(g)
 =
\frac{\Gamma(n/2)}{\pi^{n/2}}
\prod\limits_{j=1}^n
(1+[\Upsilon^{n-j}(g)]_1)^{\lambda_j}\cdot d\sigma_n(g)
.\end{equation}

\section{\hspace{-4mm}.\hspace{2mm}
EXAMPLES: CALCULATION OF SOME MATRIX INTEGRALS}

\subsection{\hspace{-5mm}.\hspace{2mm}
\bf  Some Integrals over Classical Groups.} Theorem 1.6
immediately
yields

\smallskip

{\sc Corollary 2.1}  {\it Let $\lambda_0=\mu_0=0$. Then}
 \begin{align}
&\int_{\SO(n)}\prod_{k=1}^{n}
 \det(1+[g]_{n-k+1})^{\lambda_k-\lambda_{k-1}}\,d\sigma_n(g)=
\prod_{k=1}^n   2^{\lambda_k}
\frac{\Gamma(k-1)\Gamma(\lambda_k+(k-1)/2)}
{\Gamma((k-1)/2)\Gamma(\lambda_k+k-1)};\\
&\int_{\UU(n)}\prod_{k=1}^{n}
\det(1+[g]_{n-k+1})^{\lambda_k-\lambda_{k-1}}
\overline{\det(1+[g]_{n-k+1})}^{\mu_k-\mu_{k-1}}
\,d\sigma_n(g)
=
\prod_{k=1}^n
\frac{\Gamma(k)\Gamma(k+\lambda_k+\mu_k)}
{\Gamma(k+\lambda_k)\Gamma(k+\mu_k)};\\
&\int_{\Sp(n)}\prod_{k=1}^{n}
\det(1+[g]_{n-k+1})^{\lambda_k-\lambda_{k-1}}\,d\sigma_n(g)
= \prod_{k=1}^n
\frac{\Gamma(2k)\Gamma(2k+\lambda_k+1)}
{\Gamma(2k+\lambda_k/2)\Gamma(2k+\lambda_k/2+1)}
.\end{align}
{\it The integrals are absolutely covergent under the
following conditions}

a) {\it in the real case:} $\Re \lambda_k>-(k-1)/2$,

b) {\it in the complex case:} $\Re (\lambda_k+\mu_k)>-k$

c) {\it in the quaternionian case: $\Re \lambda_k>-(2k+1)$

\noindent
for all $k$.}

\smallskip

{\sc Remark.} Let $g\in \SO(n)$. Then
$$\det(1+[g]_{n-1})=\det(1+g)/\det(1+[\Upsilon^{n-1}(g)]_1)$$
But $\Upsilon^{n-1}(g)\in \SO(1)$ and hence it equals 1. Thus
$$\det(1+[g]_{n-1})=\tfrac12\det(1+g).$$
By this reason, the integral (2.1) depend on $\lambda_1$
in a nonessential way.    \konets

\smallskip

{\sc Remark.} The absolute convergence
of the integrals (2.1)--(2.3) follows from the absolute convergence
of the integrals  (1.11), (1.12), (1.16).
Absolute convergence
of the former integrals is a simple exercise. \konets

\smallskip

The formulas (1.5), (1.17) imply
 a more general (and more artificial) statement.

\smallskip

{\sc Corollary 2.2.} {\it Let $\lambda_0=\mu_0=0,\theta_1=0$. Then}
\begin{align}
&\int_{\SO(n)}\prod_{k=1}^{n}\det(1+[g]_{n-k+1})^{\lambda_k-\lambda_{k-1}}
\prod_{k=2}^{n}(1-|[\Upsilon^{n-k}(g)]_1|^2)^{\theta_j/2}d\sigma_n(g)=
\\
&\qquad\qquad\qquad\qquad\qquad
=
\frac{\Gamma(n/2)}{\pi^{n/2}}
 \prod_{k=1}^n 2^{\lambda_k+k+\theta_k-2}
 \frac{\Gamma((k+\theta_k-1)/2)\Gamma(\lambda_k+(k+\theta_k-1)/2)}
  {\Gamma(\lambda_k+k+\theta_k-1)}\nonumber;\\
&\int_{\UU(n)}\prod_{k=1}^{n}
\det(1+[g]_{n-k+1})^{\lambda_k-\lambda_{k-1}}
\overline{\det(1+[g]_{n-k})}^{\mu_k-\mu_{k-1}}
\prod_{k=2}^{n}(1-|[\Upsilon^{n-k}(g)]_1|^2)^{\theta_j}d\sigma_n(g)
=      \\
&
\qquad\qquad\qquad\qquad\qquad\qquad\qquad
\qquad\qquad\qquad
=
(n-1)!\,
\prod_{k=1}^n \frac{\Gamma(k+\theta_k-1)\Gamma(k+\theta_k+\lambda_k+\mu_k)}
{\Gamma(k+\theta_k+\lambda_k)\Gamma(k+\theta_k+\mu_k)}\nonumber;\\
&\int_{\Sp(n)}\prod_{k=1}^{n}\det(1+[g]_{n-k+1})^{\lambda_k-\lambda_{k-1}}
\prod_{k=2}^{n}(1-|[\Upsilon^{n-k}(g)]_1|^2)^{2\theta_j}
d\sigma_n(g) =   \\
&
\qquad\qquad\qquad\qquad\qquad\qquad\qquad\qquad
= (2n-1)!
\prod_{k=1}^n
\frac{\Gamma(2(k+\theta_k-1))\Gamma(2(k+\theta_k)+\lambda_k+1)}
{\Gamma(2(k+\theta_k)+\lambda_k/2)\Gamma(2(k+\theta_k         )+\lambda_k/2+1)}  \nonumber
.\end{align}

{\sc Proof.} Consider the case $\K=\R$.
By 1.10, the integrand in (2.4) has the form
$$
\prod\limits_{j=1}^n
(1+[\Upsilon^{n-j}(g)]_1)^{\lambda_j}\cdot
            (1-|[\Upsilon^{n-k}(g)]_1|^2)^{\theta_j/2}
            $$
Thus, the integrand is a function in the variables
 $  x_j= [\Upsilon^{n-j}(g)]_1$.
Applying the formula (1.5),
we reduce the integral to the form
$$
 \frac{\Gamma(n/2)}{\pi^{n/2}}  2^{\lambda_1}
\int_{-1}^1\dots\int_{-1}^1
\prod\limits_{j=2}^n  (1+x_j)^{\lambda_j}
\prod\limits_{j=2}^n   (1-x_j^2)^{(j+\theta_j-3)/2}
\prod\limits_{j=2}^n dx_j     .
$$
This integral is reduced to (1.9).

In the cases $\K=\C,\H$, we obtain the same reduction to
the integrals (1.12), (1.16).  \konets

        \subsection{\hspace{-5mm}.\hspace{2mm}
Some Integrals over Stiefel Manifolds} Recall that a Stiefel manifold
$\Sti(m,n,\K)=\U(n,\K)/\U(n-m,\K)$ is the set of isometric embeddings
$\K^m\to\K^n$.  A projection from a group $\U(n,\K)$ to a homogeneous space
$\Sti(m,n,\K)$ is very simple: we take a unitary matrix and delete
                   its    last $n-m$  rows.

 Assume $\lambda_{0}=\dots=\lambda_{n-m}=0, \mu_{0}=\dots=\mu_{n-m}=0$
 in the integrals (2.1)--(2.3).
 Then the integrand depends only of the first $m$ rows
                        of a matrix. Therefore, we can consider the integral
                        as an integral over the Stiefel manifold
$\Sti(m,n,\K)$.

\subsection{\hspace{-5mm}.\hspace{2mm}
Some Integrals over Matrix Balls} As in Subsection 1.2, we denote
            the matrix ball by $\B_m(\K)$.
 Let  $dZ$ be the Lebesgue measure on $\B_m(\K)$
normalized as in Lemma 1.4.

\smallskip

{\sc Proposition 2.3}
\begin{align}
&
 \int_{\B_m(\R)}\det(1-Z^*Z)^{(\alpha-m-1)/2}
\prod_{k=1}^{m}\det(1+[Z]_{m-k+1})^{\lambda_k-\lambda_{k-1}}dZ=\\
&
\qquad\qquad\qquad\qquad\qquad\qquad
=c^{(m)}_\R((\alpha-m+1)/2) \prod_{k=1}^m
\frac{\Gamma(k+\alpha-1)\Gamma(\lambda_k+(\alpha+k-1)/2)}
{\Gamma((k+\alpha-1)/2)\Gamma(\lambda_k+\alpha+k-1)};\\
&
\int_{\B_m(\C)} \det(1-Z^*Z)^{\alpha-m}
\prod_{k=1}^{m}      \left\{
\det(1+[Z]_{m-k+1})^{\lambda_k-\lambda_{k-1}}
\overline{\det(1+[Z]_{m-k+1})}^{\mu_k-\mu_k}\right\}  dZ  =\\
 &
\qquad\qquad\qquad\qquad\qquad\qquad\qquad \qquad
=c^{(m)}_\C(\alpha-m+1)
\prod_{k=1}^m \frac{\Gamma(k+\alpha)\Gamma(k+\alpha+\lambda_k+\mu_k)}
{\Gamma(k+\alpha+\lambda_k)\Gamma(k+\alpha+\mu_k)} \nonumber   ;
\\
&
 \int_{\B_m(\H)} \det(1-Z^*Z)^{2(\alpha-m)+1}
\prod_{k=1}^{m}\det (1+[Z]_{m-k+1})^{\lambda_k-\lambda_{k-1}}  dZ=
\\
&
\qquad\qquad\qquad\qquad\qquad
 =
 c^{(m)}_\H(2(\alpha-m+1))
 \prod_{k=1}^m
\frac{\Gamma(2(k+\alpha))\Gamma(2(k+\alpha)+\lambda_k+1)}
{\Gamma(2(k+\alpha)+\lambda_k/2)\Gamma(2(k+\alpha)+\lambda_k/2+1)}
   \nonumber
,\end{align}
{\it where the constants $c^{(m)}_\K(\cdot)$ are the same as in Lemma} 1.4.

\smallskip

{\sc Proof} will be given in the case $\K=\R$.
Consider another parameter
$\beta=\alpha+m$. First let $\beta$ be an integer, $\beta>2m$.
Consider the integral (evaluated before, corollary 2.1).
\begin{equation}
c^{(m)}_\R((\beta-2m+1)/2)
\int_{\SO(\beta)}\prod_{k=1}^m
  \det(1+[g]_{m-k+1})^{\lambda_k-\lambda_{k-1}}
d\sigma_\beta(g)
.\end{equation}
The integrand depends only on the matrix
$[g]_m\in\B_m$. Hence, we can consider the integral as
an integral over $\B_m$.
Using Theorem 1.3,
 we convert (2.11) to the form (2.7).

Thus, the required statement
is proved for the integer values of $\alpha>m+1$.

Fix
\begin{equation}
\lambda_1\ge\lambda_2\ge\dots\ge\lambda_m.
\end{equation}
Then, for $\Re\alpha> m+1$, the left side
of the integral is a bounded holomorphic function
in $\alpha$ in the domain $\Re\alpha >m+1$.
Indeed, $\det(1-Z^*Z)\le 1$ for $Z\in\B_m$
and thus the integrand (for fixed $\lambda_j$) is bounded.

It can easily be checked that the product of $\Gamma$
also is bounded in the same domain.
By the Carlson theorem%
\footnote{If
$f(z)$ is holomorphic and bounded for $\Re z>0$
and if $f(n)=0$ for all $n=1,2,\dots $, then $f(z)=0$,
see, for instance \cite{AAR}, Theorem 2.8.1.}
the left part (2.7) and the right part  (2.8)
coincide in the whole domain $\Re\alpha> m+1$.

The analytic continuation allows to omit the condition
(2.12).
\konets

\subsection{\hspace{-5mm}.\hspace{2mm}
Some Integrals Over Spaces of Anti-Hermitian Matrices}
Let us change the variable $g=-1+2(X+1)^{-1}$ in the integral (2.1).
Obviously, $X$ is a skew-symmetric matrix.
                        Let us calculate the new integrand.
We represent $g$ as a $(m+(n-m))\times(m+(n-m))$ block matrix
$g=\matr{P&Q\\R&T}$.
 By Proposition 1.5,
                        we obtain
\begin{equation}
\det(1+[g]_m)=\det(1+P)=
  \det(1+g)\cdot\det\left(1+T-R(1+P)^{-1}Q\right)^{-1}.
\end{equation}
Further,
$$2(1+X)^{-1}=1+g=1+\matr{P&Q\\R&T}. $$
Hence, the expression $\det(1+g)$ transforms to $\const\cdot\det(1+X)^{-1}$.
The Frobenius formula (1.1) gives
\begin{equation}
2(1+X)=(1+g)^{-1}=2\matr{\dots&\dots\\ \dots&
 \bigl( 1+T-R(1+P)^{-1}Q\bigr)^{-1}}
\end{equation}
(we only write the block which is interesting for us).
Therefore, by
(2.13),(2.14), the expression $\det(1+[g]_{n-k+1})$ converts to  the form
$$\const\cdot\det(1+\{X\}_{k-1})\cdot\det(1+X)^{-1}$$
                        (according to the notation of Subsection 1.1).

The Jacobian of the transformation   $g=-1+2(X+1)^{-1}$ equals
$\const\cdot\det(1+X)^{-(n-1)/2}$ (see \cite{Hua}, \S 3.7)
and, finally, the integral transforms to the form
\begin{equation}
\const\cdot
\int
\prod_{k=2}^{n}\left(\det(1+[X]_{k-1})^{\lambda_k-\lambda_{k-1}}\right)
\det(1+X)^{-\lambda_{n}-(n-1)/2}dX,
\end{equation}
where the integration is taken over the space of real skew-symmetric
                        matrices.

In the same way, the integrals (2.2)--(2.3) transform
 into integrals over the space
of anti-Hermitian ($X=-X^*$)  matrices  over $\C$ or $\H$.
For instance, in the complex case we obtain an arbitrary
integral of the form
\begin{equation}
\int_{X+X^*=0}
\prod_{k=1}^{n}\det(1+[X]_k)^{a_k}
\det(1-[X]_k)^{b_k}dX
.\end{equation}

\subsection{\hspace{-5mm}.\hspace{2mm}
Some Integrals over Spaces of Dissipative Matrices}

Let us transform the integral (2.7)
as in Subsection 2.4. It can easily be checked that
$$\det(1-Z^*Z)=\frac{\det(2(X+X^*))}{|\det(1+X)|^2},$$
the Jacobian is given by
$$dZ=(1+X)^{-2m}dX,$$
and the integral (2.7) transforms to
\begin{equation}
\const\cdot\int\det(X+X^*)^{(\alpha-m-1)/2}
\prod_{k=1}^{m}\left(\det(1+[X]_k)^{\lambda_k-\lambda_{k-1}}\right)
\det(1+X)^{-\lambda_m-\alpha-m+1}   dX
,\end{equation}
where the integration is taken over the space of real matrices $T$
                        satisfying the condition: the matrix $T+T^*$
                        is positive definite.

The integrals (2.9)--(2.10) can be transformed in a similar way.

\smallskip

{\sc Remarks.}
a) The integral (2.17) and the Cayley transforms
of the integrals (2.9), (2.10) are partial cases of the matrix
$\B$-function introduced in \cite{Nerb}
(this $\B$-function extends Gindikin's $\B$-function,
 see \cite{Gin}, \cite{FK}).
The Cayley transform of (2.9) is also one of
Upmeier--Unterberger integrals, \cite{UU},
see also \cite{AZ}.

b) A way of separation of variables described in
\cite{Nerb} also allows to evaluate the integrals (2.15),
(2.16) but they were not evaluated in that paper.
The integral (2.16) can also be evaluated using the inverse
Laplace transform in a way explained in \cite{FK}
(but it was not evaluated in this book).
The integral (2.1) was evaluated and
the integrals (2.2)--(2.3) were announced in
\cite{Nerr}.

c) Calculations in this paper are almost verbal (except local difficulties
in 1.9),
and they provide an explanation for the existence of explicit formulas.
     \konets



\section{\hspace{-4mm}.\hspace{2mm}INVERSE LIMITS OF ORTHOGONAL
GROUPS}

In order to be concrete, we consider only the case
$\K=\R$.

\subsection{\hspace{-5mm}.\hspace{2mm}
Inverse Limit of Orthogonal Groups}
                        Consider a chain of the maps
(defined almost everythere)
\begin{equation}
\cdots\stackrel{\Upsilon^1}{\leftarrow}\OO(k)\stackrel{\Upsilon^1}
{\leftarrow}
\OO(k+1)  \stackrel{\Upsilon^1}{\leftarrow}\OO(k+2)
\stackrel{\Upsilon^1}{\leftarrow} \cdots
\end{equation}
Let us fix a sequence of real numbers
 $\lambda_1,\lambda_2,\dots$ satisfying the condition $\lambda_k>-(k-1)/2$.
Consider in each space
$\OO(k)$ a probability measure with the density
$$C(\lambda_1,\dots,\lambda_{n})^{-1}
\prod_{k=1}^{n}\det(1+[g]_{n-k+1})^{\lambda_k-\lambda_{k-1}}$$
                        with respect to the Haar measure.
 The constant $C(\cdot)$ is given by  (2.1).
By Theorem 1.6, our measures are consistent with the maps
                        $\Upsilon^1$.
Therefore, by the Kolmogorov's theorem about inverse limits
(see for instance \cite{Shi}), we obtain a canonically
defined measure on the inverse limit of the chain (3.1).
                        (This measure depends on the sequence
$\lambda_1,\lambda_2,\dots$).

We denote this inverse limit
(equipped with the probability measure)
 by $(\calo_{\lambda_1,\lambda_2,\dots},
\nu_{\lambda_1,\lambda_2,\dots})$
and we will call it the {\it virtual orthogonal group}.

We also denote by $\Upsilon^{\infty-k}$
the canonical map
$$ \Upsilon^{\infty-k}: \calo_{\lambda_1,\lambda_2,\dots}\to \SO(k).$$

This family of measures seems too large.
In Subsections 3.4--3.6 we discuss two natural
special cases.

{\sc Remark.}
Obviously, the space $\calo_{\lambda_1,\lambda_2,\dots}$
is not a projective limit in category of groups.
But an orthogonal group $\SO(n)$
is also the symmetric space $G/K=\SO(n)\times\SO(n)/\SO(n)$,
where $K$ is embedded to $G$ as the diagonal subgroup;
the maps $\Upsilon^m$ are quite natural as maps
of symmetric spaces (see \cite{NerKS}).
Hence  $\calo_{\lambda_1,\lambda_2,\dots}$
can be considered as a projective limit of symmetric spaces.
                   \konets

\subsection{\hspace{-5mm}.\hspace{2mm}
Projection of  $(\calo_{\lambda_1,\lambda_2,\dots},
\nu_{\lambda_1,\lambda_2,\dots})$ to a Cube}
Consider the product $[-1,1]^\infty$ of
segments $[-1,1]$
equipped with the product of the measures
\begin{equation}
\frac{2^{-\lambda-k+2}}{\B(\lambda_k+(k-1)/2,(k-1)/2)}
(1+x_k)^{\lambda_k}(1-x^2_k)^{(k-3)/2}
dx_k
,\end{equation}
where $k=2,3,\dots$.

We define the map
 $(\calo_{\lambda_1,\lambda_2,\dots},
\nu_{\lambda_1,\lambda_2,\dots}) \to [-1,1]^\infty $
by the formula
$$\omega\mapsto
([\Upsilon^{\infty-2}(\omega)]_1, [\Upsilon^{\infty-3}(\omega)]_1,\dots).$$
Obviously, the image of the measure
$\nu_{\lambda_1,\lambda_2,\dots}$
is our measure on the cube.

\subsection{\hspace{-5mm}.\hspace{2mm} Quasiinvariance}

Denote by   $\O(\infty)$ the group of all orthogonal operators in the
                        real Hilbert space $l_2$. Denote by
$\SO(\infty)^{\rm fin} $ the group of matrices $g\in\O(\infty)$
such that

 a) $g-1$ has only finite number of nonzero matrix elements

 b) $\det(g)=1$

It is convenient to think that {\it matrices $g\in\O(\infty)$
are infinite upwards and to the left}. We assume
that the subgroup
$\SO(k)\subset\O(\infty)$ corresponds to right lower
$k\times k$ block of infinite matrices.

Let $A,B\in\OO(k)$. Denote by $1_n$
the $n\times n$ unit matrix. Consider the map
$\Upsilon^n: \SO(n+k)\to\SO(k)$. Obviously,
$$\Upsilon^n
\left\{ \matr{1_n&0\\0&A}\matr{P&Q\\R&T} \matr{1_n&0\\0&B}  \right\}=
A\Upsilon^n \matr{P&Q\\R&T}  B.$$
This yields that for all $\cA,\cB\in \O(\infty)^{\rm fin} $
the transformation
\begin{equation}
S\mapsto \cA S\cB
\end{equation}
 of the virtual orthogonal group
  $(\calo_{\lambda_1,\lambda_2,\dots},
\nu_{\lambda_1,\lambda_2,\dots})$
                        is well defined.

\smallskip

{\sc Proposition 3.1.} {\it The measure
 $\nu_{\lambda_1,\lambda_2,\dots}$
is quasiinvariant with respect to the action of the group
$\SO(\infty)^{\rm fin}\times\SO(\infty)^{\rm fin}$.
The Radon--Nikodym derivative is given by the formula}
$$
\prod_{j=1}^\infty
\left( \frac{1+[\Upsilon^{\infty-j}(\cA S\cB)]_1}
            {1+[\Upsilon^{\infty-j}(S)]_1}\right)^{\lambda_j}
.$$

{\sc Remark.} Since $A,B\in \SO(\infty)^{\rm fin}$,
only finitely many factors of this product
differ from 1.                            \konets

\smallskip

{\sc Proof.} Let $A,B\in \SO(k)$.
Consider the transformation
\begin{equation}
S\mapsto
 \matr{1_n&0\\0&A} S  \matr{1_n&0\\0&B}
\end{equation}
of the group $\SO(n+k)$.
Its Radon--Nikodym derivative is a ratio of densities, and it
is equal to
\begin{equation}
\prod_{j=1}^{n+k}
\Biggl(\frac{1+\bigl[\Upsilon^{n+k-j}
\bigl(\bigl(\begin{smallmatrix}1_n&0\\0&A\end{smallmatrix}\bigr)
S \bigl(
\begin{smallmatrix}1_n&0\\0&B \end{smallmatrix}\bigr)\bigr)\bigr]_1}
            {1+[\Upsilon^{n+k-j}(S)]_1}\Biggr)^{\lambda_j}
.\end{equation}
By Lemma 1.1,  for $j>k$
$$\Upsilon^{n+k-j}
 \bigl(\bigl(\begin{smallmatrix}1_n&0\\0&A\end{smallmatrix}\bigr)
 S
\bigl( \begin{smallmatrix}1_n&0\\0&B \end{smallmatrix}\bigr)\bigr)
=
 \bigl(\begin{smallmatrix}1_{j-k}&0\\0&A\end{smallmatrix}\bigr)
\Upsilon^{n+k-j}( S )
\bigl( \begin{smallmatrix}1_{j-k}&0\\0&B \end{smallmatrix}\bigr)
.$$
Thus, for $j>k$ we have
$$[\Upsilon^{n+k-j}
  \bigl(\bigl(\begin{smallmatrix}1_n&0\\0&A\end{smallmatrix}\bigr)
  S
 \bigl( \begin{smallmatrix}1_n&0\\0&B \end{smallmatrix}\bigr)\bigr)
]_1=[\Upsilon^{n+k-j}(S)]_1$$
and, hence, the product (3.5) is reduced to
$\prod_{j=1}^k$.
Thus, the Radon--Nikodym derivative of the transformation
(3.4) depends only on $\Upsilon^{n+k-k}(S)$.
 Hence, the Radon--Nikodym derivatives of the maps
(3.4) (where $A,B\in \SO(k)$ are fixed) form a compatible system
of functions
with respect to the chain (3.1). This implies both statements.
                                                 \konets

\subsection{\hspace{-5mm}.\hspace{2mm}
Hua--Pickrell measures}
 Let $\lambda>-1/2$. Consider the probability
                        measure on $\OO(n)$ given by the formula
 $$    \nu^n_\lambda=C(n,\lambda)\det(1+g)^\lambda,$$
 where  $C(n,\lambda)$ is a constant.

This corresponds to the case
$$\lambda_1=\lambda_2=\dots=\lambda$$
in the construction of Subsection 3.1.
Let us denote the inverse limits of the  measure spaces
$(\OO(n),\nu^n_\lambda)$
by $({\calo}_\lambda(\infty),\nu_\lambda)$.
We call   $\nu_\lambda$ by Hua--Pickrell measure.

\smallskip

{\sc Proposition 3.2.}{\it {\rm a)}
 The measure $\nu_\lambda$ on ${\calo}_\lambda(\infty)$
 is quasi invariant {\rm(} in the case $\lambda=0$ it is invariant{\rm)}
 with respect to
the action of the group
$\SO(\infty)^{\rm fin} \times \SO(\infty)^{\rm fin} $. Moreover,
for $A,B\in \SO(k)\subset\SO(\infty)$
the Radon-Nikodym derivative of the transformation
$S\mapsto ASB$ is equal to
$$\Bigl[ \frac{\det(1+A\Upsilon^{\infty-k}(x)B)}
             {\det(1+\Upsilon^{\infty-k}(x))}
\Bigr]^\lambda
.$$

{\rm b)} The diagonal action
$S\mapsto A^{-1}SA$ of the group $\SO(\infty)^{\rm fin} $
                        extends to an  invariant action of the group
$\O(\infty)$.}

\smallskip

{\sc Proof.} a) Let $S=\begin{pmatrix}P&Q\\R&T\end{pmatrix}\in\SO(n+k)$.
Then the Radon-Nikodym derivative of the transformation
$$S\mapsto
 \matr{1_n&0\\0&A} S \matr{1_n&0\\0&B}$$
is
\begin{multline*}   \frac{
\det\left\{\matr{1_n&0\\0&1_k}+
    \matr{1_n&0\\0&A}\matr{P&Q\\R&T} \matr{1_n&0\\0&B}  \right\}^{\lambda}}
{\det\left\{\matr{1_n&0\\0&1_k}+
   \matr{P&Q\\R&T}   \right\}^{\lambda}}=\\
=\frac{\det(1+P)^{\lambda}\det(1+ATB-AR(1+P)^{-1}QB)^{\lambda}}
{\det(1+P)^{{\lambda}}\det(1+T-R(1+P)^{-1}Q)^{{\lambda}}     }
=
\frac{\det(1+A\Upsilon^n(S)B)^{\lambda}}
{\det(1+\Upsilon^n(S))^{\lambda}}
.\end{multline*}
The Radon--Nikodym derivative depends only on
$\Upsilon^n(S)\in\SO(k)$ and this implies a).
Clearly, a) is also a corollary of the Proposition 3.1.

b) This statement is very simple but its proof uses some
technique. By a criterion from \cite{Ols1}, the representation of
the diagonal group
$\O(\infty)^{\rm fin}$ in $L^2(\calo_\lambda)$ is weakly continuous.
Hence, it extends to the group
$\O(\infty)$. The group $\O(\infty)$
acts by measure preserving transformations
and, hence, the group $\O(\infty)$ acts by
polymorphisms (see \cite{Nerbook}, chapter 8).
But an invertible polymorphism is a measure preserving
transformation.                                         \konets

\smallskip

{\sc Remark.}
We see that the group of symmetries of
the space $(\calo_\lambda, \nu_\lambda)$
is larger than that for a general space
 $(\calo_{\lambda_1,\lambda_2,\dots},
\nu_{\lambda_1,\lambda_2,\dots})$.
\konets

\smallskip

{\sc Remark.} Consider the group
$\O(\infty)\times\O(\infty)$ and its subgroup $G$
that consists of pairs
$(g_1,g_2)\in\O(\infty)\times\O(\infty)$
such that $g_1 g_2^{-1}$
is a Hilbert--Schmidt operator (this is one of
Olshanski's $(G,K)$-pairs, see \cite{Ols1}, \cite{Nerbook}.
 It is natural
to think
that our action extends to a quasiinvariant action
of $G$.
\konets

\smallskip

{\sc Remark.} Our construction for $\lambda=0$
is the Shimomura construction, \cite{Shim}.
 D.Pickrell (\cite{Pic}) constructed a 1-parametric family
of probability measures on inverse limits of Grassmannian
                        $\UU(2n)/\UU(n)\times\UU(n)$.
G.I.Olshanski(\cite{Ols}) observed that a Pickrell's type construction
extends to all 10 series of classical compact symmetric spaces
(this can be observed from Hua Loo Keng calculations\footnote{%
In Theorem 2.2.2 of \cite{Hua} Hua exactly claims a projectivity
of some system of measures (for noncompact symmetric spaces).}
\cite{Hua}, chapter 2). In particular, it can be carried out for
              the          classical groups
$\UU(n),\SO(n),\Sp(n)$. Our construction of the measure
$\nu_\lambda$ is equivalent to this construction
(in the complex case our construction
 gives an additional parameter).

\smallskip

\subsection{\hspace{-5mm}.\hspace{2mm}
Some Integrals over the space $({\calo}(\infty),\nu_\lambda)$}

Consider the space $({\calo}_\lambda(\infty),\nu^\lambda)$
equipped with Hua--Pickrell measure.
The construction of Subsection 3.1 can be considered
as a construction of a large family of functions on
the space
 $({\calo}_\lambda(\infty),\nu^\lambda)$
 with explicitly computable integrals.

 Let $x_j$ be the coordinates on the cube as in 3.2.

\smallskip

{\sc Proposition 3.3.} {\it
Consider a sequence $\lambda_1,\lambda_2,\dots\in\C$ such that
$\sum |\lambda_k-\lambda|<\infty$, $\lambda_k>-(k-1)/2$.
Let us define  the function
 $\Phi\{\lambda_1,\lambda_2,\dots\}(\omega)$ on ${\calo}(\infty)$
                        by the formula
\begin{multline}                  \Phi\{\lambda_1,\lambda_2,\dots\}(x)  =
2^{\lambda_1-\lambda}
\prod_{j=2}^\infty\det(1+x_j)^{\lambda_{j}-\lambda}
=
\prod_{k=1}^\infty
(1+[\Upsilon^{\infty-k}
    (\omega)]_1)^{\lambda_k-\lambda}
= \\=
\lim_{k\to\infty}
\frac{\prod_{j=1}^{k}
 \det(1+[\Upsilon^{\infty-k}(\omega)]_j)^{\lambda_{j}-\lambda_{j-1}}}
{\det(1+\Upsilon^{\infty-k}(\omega))^\lambda}
  .\end{multline}
Then the limit exists almost everywhere on
 $({\calo}(\infty),\nu^\lambda)$ and}
\begin{equation}
\int_{{\calo}_{\cal \lambda}(\infty)}
         \Phi\{\lambda_1,\lambda_2,\dots\}\,d\nu^\lambda=
\prod_{k=1}^\infty
2^{\lambda_k-\lambda}
\frac
   {\Gamma(\lambda_k+(k-1)/2)\Gamma(\lambda+k-1)}
   {\Gamma(\lambda+(k-1)/2)\Gamma(\lambda_k+k-1)}
.\end{equation}

{\sc Proof.} Let us transform the expression (3.6) to the form
$$\prod_{k=1}^\infty
  (1+[\Upsilon^{\infty-k}
      (\omega)]_1)^{\lambda_k-\lambda}
.$$
First, let us  prove existence of the functions  $\Phi$.
It is sufficient to prove the  convergence
\begin{equation}
\prod_{k=2}^\infty(1+x_k)^{\lambda_k-\lambda}
\end{equation}
on the cube $[-1,1]^\infty$ equipped with measure
(3.2). This is equivalent to the  convergence
of the series
\begin{equation}
\sum_{k=2}^\infty(\lambda_k-\lambda)\ln(1+x_k)
.\end{equation}
 By the Kolmogorov--Khintchin
theorem on series of independent random variables
(see \cite{Shi}),
it is sufficient to prove the absolute convergence of
the series of means and the convergence of the series of variances, i.e.,
\begin{align}
&\sum C_k^{-1}\Bigl|(\lambda_k-\lambda)\int_{-1}^1
  \ln(1+x)(1+x)^\lambda(1-x^2)^{(k-3)/2}dx\Bigr|<\infty;\\
& \sum C_k^{-1}|\lambda_k-\lambda|^2\int_{-1}^1
\ln^2(1+x)(1+x)^\lambda(1-x^2)^{(k-3)/2}dx<\infty
,\end{align}
where
$$C_k=\int_{-1}^1 (1+x)^\lambda(1-x^2)^{(k-3)/2}dx.$$
The Laplace method gives the asymptotics
$C_k=\const\cdot k^{-1/2}(1+o(1))$
and $\const\cdot  k^{-3/2}(1+o(1))$
for the integrals under the sums (3.10), (3.11).
This implies the a.s. convergence of (3.9).

The product (3.8) is dominated by
\begin{equation}
\prod_{k=2}^\infty\max\left((1+x_k)^{\lambda_k-\lambda},
      (1+x_k)^{\lambda-\lambda_k}\right)
.\end{equation}
By the Lebesgue's theorem on dominated convergence,
it is sufficient to prove that the last expression
is integrable. The integral of (3.12) is
$$
\prod_{k=2}^\infty C_k^{-1}\Bigl(\int_0^1
(1+x_k)^{|\lambda_k-\lambda|}(1+x)^\lambda(1-x^2)^{(k-3)/2}dx
+
\int_{-1}^0
(1+x_k)^{-|\lambda-\lambda_k|}(1+x)^\lambda(1-x^2)^{(k-3)/2}dx
\Bigr)
.$$
The Laplace method gives the asymptotics
$\const\cdot k^{-1/2}|\lambda_k-\lambda|(1+o(1))$
for the integrals under the product, and this implies the required
statement.               \konets

{\sc Remark.} Author thinks that condition
$\sum|\lambda_k-\lambda|<\infty$
is not necessary. \konets

\subsection{\hspace{-5mm}.\hspace{2mm}
Measures on Inverse Limits of Stiefel Manifolds}
Let us fix $p>0$.
We denote by $\psi_k(g)$
the function
$$\det(1+[g]_{k})^\lambda.$$
 on $\OO(k+p)$.
 Obviously, the function  $\psi_k(g)$ 
is invariant with respect to the action of the group
 $\SO(p)$ given by the formula
$$g\mapsto \matr{1&0\\0&A} g;\qquad g\in\SO(k+p),A\in \SO(p)$$
Hence, we can consider the function  $\psi_k(g)$
as a function on the Stiefel manifold (see Subsection 2.2)
$\Sti(k+p,k)$.
Denote by  $\nu^\lambda_k$ the probability measure on
$\Sti(k+p,k)$ with the density $\const\cdot\psi_k(g)$.
The projections from the chain (3.1) commute with the action of
$\SO(p)$. Therefore, we can consider a chain of quotient-spaces
$\Sti(k,k+p)=\SO(k+p)/\SO(p)$ equipped with the measures
$\nu^\lambda_k$ :
$$\cdots\stackrel{\Upsilon^1}{\leftarrow}\Sti(k,k+p)
\stackrel{\Upsilon^1}{\leftarrow}
\Sti(k+1,k+p+1)  \stackrel{\Upsilon^1}{\leftarrow}\Sti(k+2,k+p+2)
\stackrel{\Upsilon^1}{\leftarrow} \cdots                         $$
We denote by $(\Sti(\infty,\infty+p),\nu_\lambda)$
 the inverse limit of this chain.
Denote by
$\nu^\lambda$ the canonical measure on this limit.

We define the group
$\O(\infty)^{\rm fin} \times \O(\infty+p)^{\rm fin} $
as  the inductive limit of the groups
$\O(k)\times\O(k+p)$ as $k\to\infty$.

{\sc Proposition 3.4} a) {\it  The measure $\nu^\lambda$  is quasi invariant
with respect to the action
of the group $\O(\infty)^{\rm fin} \times \O(\infty+p)^{\rm fin} $

\rm b) \it The action $A:S\mapsto A S\matr{A&\\&1_k}$ of the diagonal group
$\O(\infty)^{\rm fin}$
on $(\Sti(\infty,\infty+p),\nu^\lambda)$
extends to an invariant action of the group $\O(\infty)$.}

\end{document}